\documentclass[twocolumn,twoside,letterpaper,11pt]{article}

\usepackage{html}
\usepackage{graphicx,xrrc,natbib,html}

\usepackage{amssymb}

\def\chandra{{\it CHANDRA}}
\def\exosat{{\it EXOSAT}}
\def\ec{$\eta$ Carinae}
\def\rxte{{\it RXTE}}
\def\asca{{\it ASCA}}
\def\rosat{{\it ROSAT}}

\def\xmm{{\it XMM-Newton}}

\def\sax{{\it SAX}}
\def\vterm{\mbox{$V_{\infty}$}}
\def\mdot{\mbox{$\dot{M}$}}
\def\mdunit{\mbox{$M_{\odot}$ yr$^{-1}$}}
\def\ks{~km s$^{-1}$}
\def\apj{ApJ}
\def\aj{AJ}
\def\apjs{ApJS}
\def\apjl{ApJL}
\def\mnras{MNRAS}
\def\aap{A\&A}

\begin{document}


\title{MULTI-WAVELENGTH OBSERVATIONS OF COLLIDING STELLAR WINDS}


%
%
%
%


\author{    M. F. Corcoran                        } 
\institute{ USRA \& NASA-GSFC/LHEA         } 
\address{   Code 662, GSFC, Greenbelt, MD 21114, U.S.A.        } 
\email{     corcoran@lheapop.gsfc.nasa.gov                         } 

\author{   J. M. Pittard, I. R. Stevens, D. B. Henley, A. M. T. Pollock}
\email{jmp@ast.leeds.ac.uk, irs@star.sr.bham.ac.uk, dbh@star.sr.bham.ac.uk, apollock@xmm.vilspa.esa.es}


\maketitle

\abstract{Massive colliding wind binary stars serve as laboratories for the study of strong-shock physics. In these systems fundamental flow parameters (velocities, densities, directions) are often well known, and photon fields important for Compton scattering and ionization are usually defined or constrained as well. In this paper we review observed X-ray and radio characteristics of some important colliding wind systems.  We discuss how X-ray and radio observations of these systems can be used to test models of the collision of massive flows in the presence of intense UV radiation and magnetic fields.  }

\section{Introduction}
\label{Introduction}

Though massive stars (stars with $M\gtrsim 8M_{\odot}$) are relatively rare objects they have a cosmic impact which is disproportionate to their numbers.  They are chemical factories, they explode and they produce massive compact objects (neutron stars and black holes).  Recent work suggests that the first stars which formed after the big bang had masses of tens or even hundreds of solar masses \citep{abel98} and that massive stars may be the progenitors of at least some types of gamma-ray bursts   \citep{heger03}. Over the last 2 decades the theoretical understanding of these stars has greatly advanced.  However testing these theories is a challenge since poorly-constrained quantities like convective core overshooting, mixing, weak-line opacity in the interior and  ionization balances, opacities, instabilities and non-smooth density distributions (``clumping'') in the atmosphere play key roles in determining stellar structure and evolution.  

Historically observations of binary stars have played an important role in testing models of stellar evolution, interiors and atmospheres, and in testing Newtonian \& non-Newtonian gravity as well.  In this paper we expand this role by discussing how binary stars can help test models of massive outflows and strong shocks.  We discuss the symbiotic way in which radio and X-ray observations of colliding wind binary systems help constrain shock models, and how, by developing better understanding of colliding wind shock physics, we can develop a deeper understanding of massive star astrophysics.  In \S \ref{windchar}, we discuss the properties of the stellar wind ``laboratory'', reviewing typical properties of winds in different settings.  In \S \ref{interactions} we discuss the various types of interactions that can be usefully observed and modelled.  We discuss the fundamental physics of colliding stellar winds in \S \ref{physics}, and show how they are applied to an important system, the extreme star \ec, in \S \ref{ec}, and to WR 140, an important WR+O binary, in \S \ref{wr140}. In \S \ref{winds} we consider how studies of colliding wind systems can help improve understanding of radiatively-driven mass loss. We discuss direct imaging of colliding wind shocks in \S \ref{imaging}, and how colliding wind emission can help define our understanding of the binary fraction in \S \ref{binaries}. Our conclusions and suggestions for future work are given in \S \ref{conclusions}.

\section{Wind Characteristics of Massive Stars}
\label{windchar}

Stars with masses of $\gtrsim 8M_{\odot}$ possess sufficient radiative flux to produce dense mass outflows from their photospheres by conversion of the outwardly-directed photon momentum to gas momentum by line absorption \citep{morton, cak}.  These radiatively-driven stellar winds produce characteristic ``P-Cygni''-type profiles in resonant transitions in the ultraviolet, in which outflowing material in front of the disk of the star produces blueshifted absorption, while material above and below the stellar disk produces predominantly redshifted emission.  Stellar winds are observationally characterized by three parameters: 

\begin{itemize}
\item the terminal velocity \vterm\ which represents the maximum outflow velocity of wind material;

\item  the mass loss rate \mdot\ which measures the divergence of material flowing out from the star and which is related to the mass density $\rho$  by the conservation equation, \mdot$= 4 \pi r^{2} V(r) \rho$, where $V$ is the wind velocity at any given radius $r$;

\item and the wind acceleration parameter $\beta$, which relates the velocity at a given radius to \vterm\ as $V(r)/$\vterm $= (1-R/r)^{\beta}$ where $R$ is the stellar radius.  
\end{itemize}

In many instances and throughout this paper it is assumed that $\beta=1$.   The maximum blue velocity in saturated  absorption components is an indication of the maximum speed of the outflowing stellar wind.  Mass loss rates can be derived by detailed synthesis of the P-Cygni features since the density of the overlying wind determines both the optical depth of the P-Cygni absorption and also the strength of the emission component.  However such fine analysis requires knowledge of the ionization balance through the wind, which is often poorly constrained. Mass loss rates can be determined more directly from observations of free-free emission in the radio, since  the free-free brightness  $S_{\nu}$ in Jy is related to the mass-loss rate as $S_{\nu} \propto (\mdot/\vterm)^{4/3} D^{-2}$ \citep{wb}, where $D$ is the distance to the source, assuming the wind is spherically symmetric and smooth (assumptions which are not well justified).  For a review of wind properties from a large selection of OB stars derived from UV and radio analyses, see \citet{hp89} and \citet{bac89}.

Winds evolve as stars evolve.   Typically an OB main sequence star has a mass loss rate in the range $10^{-7}\le \mdot \le 10^{-6}$ $M_{\odot}$ yr$^{-1}$ and wind speeds $\sim 1000-2000$\ks.  Theoretically $\vterm\sim 3 V_{esc}^{1.2}$ \citep{fa86} where $V_{esc}$ is the escape speed at the stellar photosphere.  As the star evolves the photosphere tends to expand, lowering the escape speed and thus lowering the observed wind speed, so that, in giant OB stars, the wind speeds are typically less than the winds of dwarf stars of similar spectral type.  Because it's easier to escape the photosphere of a giant star, the mass loss rates of giants are larger than those of dwarfs, typically by a factor of a few.  Very massive stars which enter the Wolf-Rayet phase exhibit very large mass loss rates ($\sim$ few $\times 10^{-5}$ \mdunit) with rather high terminal velocities.  The most massive of these stars may enter the so-called ``Luminous Blue Variable'' (LBV) phase in which the star is thought to be near the Eddington stability limit. In this phase the star may experience episodes of extreme mass loss ($\sim$ few $\times 10^{-4}-10^{-3}$ \mdunit) while the wind terminal velocities drop to a few hundred \ks.

\section{Types of Interactions}
\label{interactions}

The wind from a massive star will eventually collide with something.   A parcel of wind material can interact with another clump or stream in the wind. Stellar winds will collide with circumstellar material to form an evacuated region or bubble; this material may be primordial interstellar material or it might be material which was ejected from the star in an earlier evolutionary phase.  In binary systems, the wind from one star will collide with the wind or surface of the companion.  Some or all of these interactions can happen concurrently. In the following we briefly discuss the most common types of interactions which are known to give rise to observable X-ray and/or radio emission in massive stars. 

\subsection{Self-Colliding Winds}
Radiatively-driven winds are inherently unstable to doppler perturbation \citep{ls70}, since a small increase in the speed of a parcel of wind will move the parcel out of the shadow of the underlying wind material, producing an increase in the acceleration as the parcel sees more of the photospheric flux.  This mechanism is thought to produce small clumps of shocked gas distributed throughout the winds of OB stars. These distributed shocks are thought to be responsible for the X-ray emission observed from single OB stars \citep{berg97}.  X-ray emission observed from OB stars is typically found to have $T\sim10^{6}$ K (suggesting that the shocks are not very strong) with little time dependence.  Recent high-resolution X-ray spectra obtained by \chandra\ and \xmm\ suggest that some of this emission arises about 2 stellar radii above the photosphere, or perhaps interior to this \citep{cass01,kahn01}.  Such shocks may also produce observable non-thermal radio emission from single massive stars.

\subsection{Superbubbles}
Since massive stars evolve quickly, most do not have time to move very far from the nebulae out of which they formed.  Most massive stars are found in HII regions and in clusters and OB associations.  These stars will directly interact with this rather dense environment by the transfer of wind kinetic energy into the circumstellar medium.  This process can produce a wind-blown ``superbubble'' \citep{jc75,jd77,rw77} filled with hot shocked gas from the thermalization of the wind kinetic energy.  Such bubbles may be observable at X-ray and radio wavelengths.  Since the wind properties (wind speeds, mass-loss rates) vary as a massive star evolves, the circumstellar region will in general be shaped by these various types of outflows.  For example, the fast thin wind of a main sequence star may carve out a large cavity within the primordial interstellar medium.  If the star enters the LBV phase its massive, slow wind will flow into this cavity largely unimpeded.  During the subsequent evolution of the star into the Wolf-Rayet phase, a less massive but faster wind develops which can eventually overtake and collide with the slower wind from the earlier phase.  All these collisions can give rise to observable X-ray and radio emission, though important emission properties (like abundances in the shocked material, the Mach number, the proportion of thermal to non-thermal emission, etc.) will generally be different at the different shock boundaries. 

\subsection{Colliding Wind Binaries}
Perhaps the most interesting type of interaction occurs in massive binaries, where the wind from the primary star (here defined to mean the star with the stronger wind) can directly collide with the wind (or, in extreme cases, the photosphere) of the secondary star.  This collision forms a ``bow shock'' around the secondary.  Again thermalization of the kinetic energy of the wind means that portions of the bow shock can be heated to temperatures of millions of degrees and can radiate strongly at X-ray energies.  This emission can be observable against the background emission from the ``distributed shocks'' in the stellar winds themselves, since the colliding wind shock is expected to have temperatures about a factor of 10 or so higher than the distributed shock models if the wind-wind collision occurs where the winds have already reached their maximum velocities.  In addition this emission should be highly variable due to intrinsic changes in the emitting region, or to changes in the amount of intervening material between the observer and the shock, or to some combination.  Along with X-ray emission these sources can produce significant particle acceleration via acceleration of free electrons by the first order Fermi process.  In the presence of a magnetic field these electrons will produce non-thermal synchrotron emission in the radio.  This population of accelerated particles could inverse-Compton scatter UV photons to higher energies, producing non-thermal X-ray or even gamma-ray emission \citep{eich93}.

\section{Simple Colliding Wind Physics}
\label{physics}

\begin{figure}
  \begin{center}
    \includegraphics[width=\columnwidth]{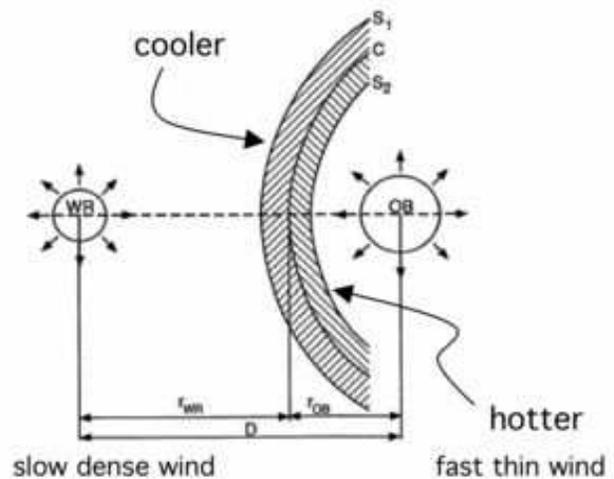}
    \caption{\small Cartoon of a colliding wind shock, adapted from \citet{ps76}. }
    \label{fig:cartoon}
  \end{center}
\end{figure}

Because the relative wind speeds are $>1000$\ks, corresponding to a thermalization energy $\sim1$ keV, the shocked gas in a wind-wind collision will produce thermal X-radiation.  The underlying physics which describes X-ray emission from wind-wind collisions was first discussed by \citet{cher67} and by \citet{ps76} with additional work by \citet{luo90}, \citet{sbp}, \citet{usov92}, and more recently by \citet{jmp97}, \citet{rw98}, and \citet{dh03}.  Though the shock structure can be complex and altered by numerous instabilities (see the paper by Julian Pittard at this meeting) the underlying physical model is actually fairly simple.  Figure \ref{fig:cartoon} shows a cartoon of the situation, adapted from  \citet{ps76}.  Consider  a massive binary composed of 2 stars, star A and star B.  In this system the dominant wind is determined by the value of the wind force ratio $\eta = \dot{p}_{A}/\dot{p}_{B} = (\mdot_{A} \vterm_{A})/(\mdot_{B} \vterm_{B})$. If $\eta>1$, the ``bow shock'' will be concave around star B; if it's less than 1, it will be concave around star A; and if $\eta=1$, the shock, in an idealized case in the absence of Coriolis forces,  forms a plane midway between the two stars. The distance $r_{s}$ from the center of star A to the bow shock, along the line of centers, is given by $r_{s}=D\sqrt{\eta}/(1+\sqrt{\eta})$ where $D$ is the separation between the two stars.  If the shocked gas is adiabatic and the winds are at their terminal velocities, then the intrinsic X-ray luminosity of the shocked gas is $L_{x} \propto \mdot^{2}/D$.  Phase dependent variations in the intrinsic X-ray luminosity will occur since the intrinsic luminosity depends on the separation between the stars, which varies with phase as long as the orbital eccentricity is $\ne 0$.  The intrinsic X-ray emission from the shock is thus greatest at periastron and is at minimum at apastron.  The observed luminosity is modified by any intervening absorption, $L_{x, obs} \propto L_{x}e^{-\sigma N_{H}}$ where $N_{H}$ is the column depth to the X-ray source and $\sigma$ the appropriate weighted atomic cross-section.  In general the absorbing material consists of interstellar medium absorption plus absorption in the wind of the frontmost star.  The component of wind absorption will generally vary as the stars revolve in their orbit. Thus, even when $e=0$ and the intrinsic X-ray luminosity is constant through the orbit, phase dependence in the observed emission can still be produced by phase-dependent changes in the wind absorption component for suitably high inclinations. 

Colliding wind systems also produce radio emission by a variety of mechanisms.  Particle acceleration by first-order Fermi processes at the shock boundary can give rise to non-thermal synchrotron emission in the presence of magnetic fields near the shock boundary \citep{sean03}. Thus the presence of non-thermal radio emission may be an indicator of the presence of a wind-wind collision.  Since the
ionized stellar winds themselves are sources of thermal free-free emission,  disentangling the contributions of thermal and non-thermal processes can be difficult.  In addition observation of the non-thermal emission depends sensitively on the radio optical depth through the wind of the frontmost star, which is usually large for stars with large mass loss rates.  In such systems the non-thermal component may well remain buried within the stellar winds, and unobservable.  


\section{A Non-Canonical Example of Colliding Winds: \ec}
\label{ec}

The enigmatic massive star \ec\ shows emission from various types of shock processes, from ejecta-ISM interactions far from the star to variations thought to be associated with a wind-wind collision near the star.  \ec\ is extremely luminous and believed to lie very near the Eddington Limit.  The star has undergone at least two major eruptions since the 1840's.   In addition to these ejection events (where the mass loss rate may have been $0.1$ \mdunit\ or larger) the star presently loses mass at a rate of about $10^{-4}-10^{-3}$\mdunit\  and a velocity of $V_{\infty}\approx 500$\ks (with perhaps higher velocities and larger mass loss rates near the stellar poles).  The star is thus a laboratory to study the interactions of these ejections with the ISM, and the interactions of the wind with the ejections.  

\begin{figure}
  \begin{center}
    \includegraphics[width=\columnwidth]{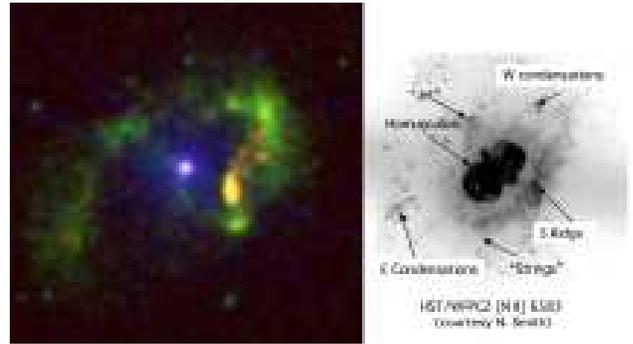}
    \caption{\small \textit{Left}: \chandra\ X-ray image of \ec.  This image was constructed from 5-100 ksec zeroth-order High Energy Transmission Grating Spectrometer observations. The colors represent different X-ray photon energies. Red: $0.1-0.5$ keV; green: $0.5-1.0$ keV; blue $1.0-10.0$ keV.  The star appears as a white point source at the center of the image. Sharp linear features are either the dispersed spectra or instrumental readout strips.  \textit{Right}: HST/WFPC2 [NII] 6583\AA\ image showing the nebulosity around \ec; the star itself can't be seen in this image but is near the middle of the Homunculus. The scale and orientation of the optical and X-ray images are the same.}
    \label{fig:ecimage}
  \end{center}
\end{figure}

The  eruption of \ec\  in 1843 was associated with a large ejection of $\gtrsim 2 M_{\odot}$ which now forms the bipolar ``Homunculus'' nebula around the star (see Figure \ref{fig:ecimage}).  Although optical spectra recently found evidence of a high-velocity structure around the Homunculus which may be indicative of the forward shock just beyond the bipolar lobes \citep{dor03}, no X-ray emission appears to be associated with this shock.   Beyond the Homunculus lies the ``outer debris field'' identified as an incomplete optical shell which may be the remains of an earlier eruption \citep{nolan78} or decelerated fast material associated with the 1843 event.  The outer debris field is associated with a ``ring'' of X-ray emitting material at the forward shock produced as the ejecta plows into the ISM \citep{chleb84, kw04}.  This outer shock is characterized by a temperature near $2\times 10^{6}$ K and absorption consistent with the ISM column as measured in the directions of neighboring stars.  Figure \ref{fig:ecimage} compares the X-ray and optical image of the nebulosities surrounding \ec.

\subsection{Radio \& X-ray Variations - Signs of a Companion Star?}

Recent evidence suggests strongly that \ec\ is in reality a binary system with a 5.53-year period.  Radio and X-ray observations in 1992 and 1993 showed what, at the time, was described as a ``brightening'' or ``outburst'' of the star \citep{dunc95,corc95}.  
This episode was associated with weakening of He I 10830\AA\ emission and an abrupt change in the infrared, events which where suggested to be periodic by \citet{pw94} and \citet{dam96}.  Continued monitoring at optical, radio and X-ray wavelengths showed that the emission over these wavebands is strongly correlated.  Every 5.5 years the radio, IR, optical and X-ray emission all  experience a brief minimum in intensity.

\begin{figure}
  \begin{center}
    \includegraphics[width=\columnwidth]{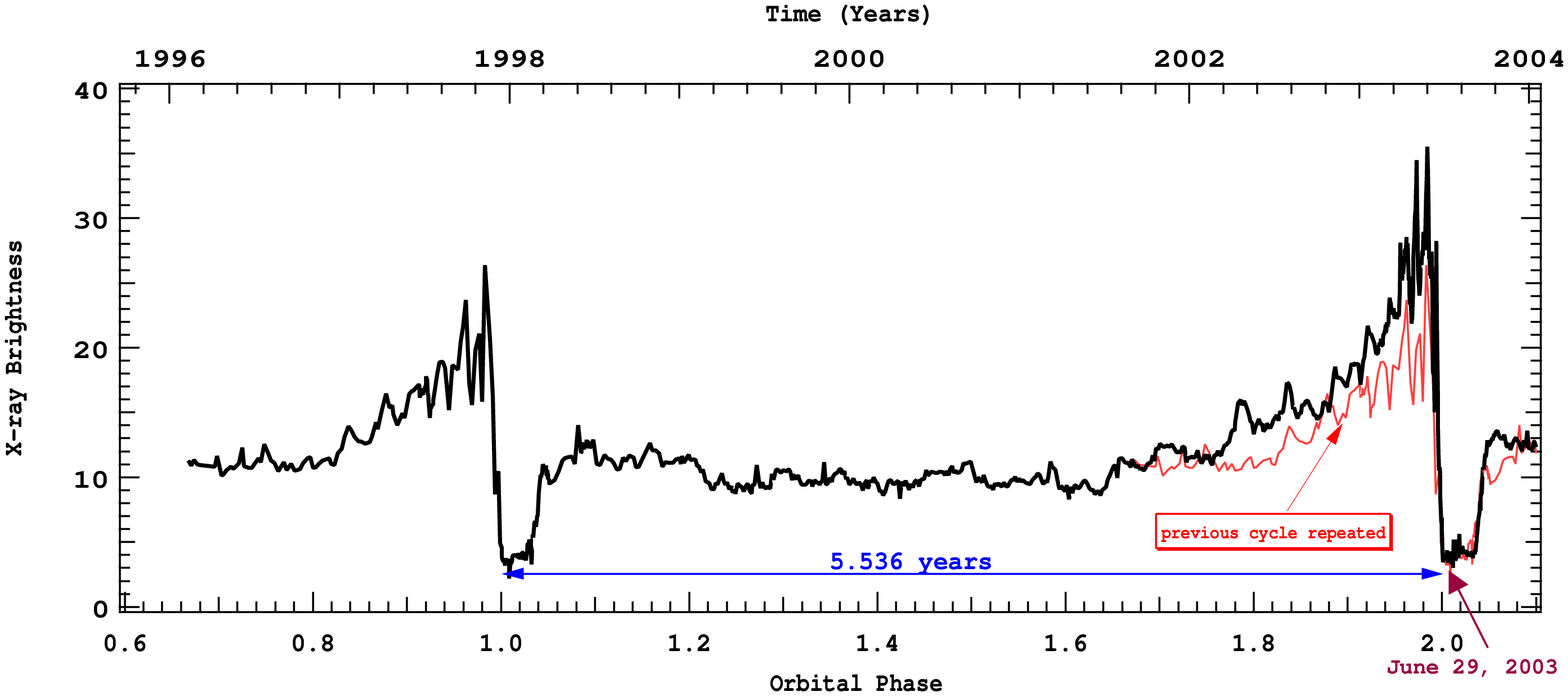}
    \caption{\small $2-10$ keV X-ray brightness of \ec\ as measured by the \rxte\ satellite from 1996-2003.  The X-ray ``lightcurve'' shows a gradual increase in X-ray brightness prior to the decline to a minimum which lasts 3 months. }
    \label{fig:lc}
  \end{center}
\end{figure}

This multiwavelength variability has been interpreted as arising from the presence of a hidden, hotter companion.  UV flux from this companion is thought to play a significant role in modulating the excitation and ionization of the nebular gas, which gives rise to variations seen in the He I 10830 line and in the free-free radio emission.  By contrast the X-ray emission is believed to arise from the wind-wind collision and is thus thought to be a direct tracer of the conditions in the  colliding wind shock.  Figure \ref{fig:lc} shows the observed X-ray brightness in the $2-10$ keV band since 1996.  The start of the X-ray minimum is given by 

\begin{center}
\begin{equation}
\mbox{UT X-ray minimum, years} = 1997.95 + 5.536E
\end{equation}
\end{center}

where the epoch is derived from daily monitoring observations with \rxte\ prior to the 1998 X-ray minimum, and the period, $P=5.526$ years,  is the interval between the start of the consecutive minima.  The nominal uncertainty on the period is less than one day.  

\begin{figure}
  \begin{center}
    \includegraphics[width=\columnwidth]{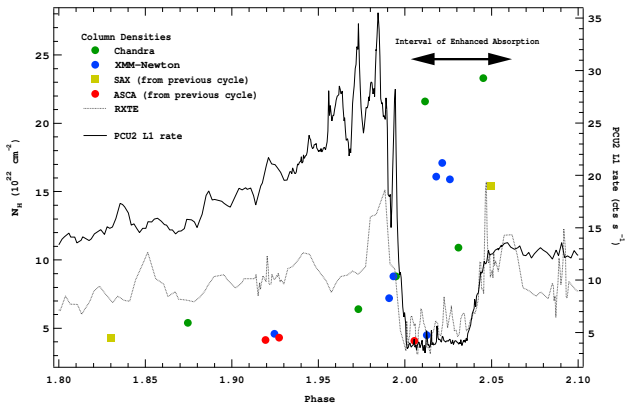}
    \caption{\small Column densities derived from fitting the X-ray spectra obtained by \asca, \sax, \chandra, \& \xmm.  The column in the interval just after the start of X-ray eclipse is markedly higher than the column prior to the eclipse. }
    \label{fig:nh}
  \end{center}
\end{figure}

Figure \ref{fig:nh} shows the column densities, $N_{H}$, derived by fitting spatially-resolved X-ray spectra obtained by  \asca, \sax, \chandra, \& \xmm\ around the time of the X-ray minima in 1998 \& 2003.  The column densities are clearly larger during the minimum than prior to the minimum.  In addition the interval of enhanced absorption extends beyond the ``recovery'' of the X-ray flux.

\begin{figure}
  \begin{center}
    \includegraphics[width=\columnwidth]{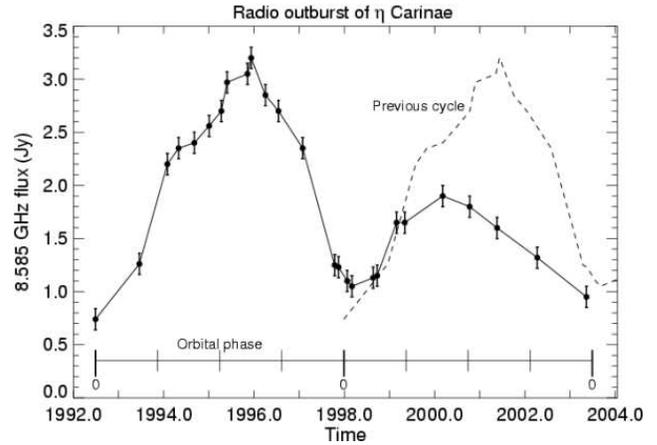}
    \caption{\small The radio lightcurve of \ec\ from Duncan \& White from 1992--2004 (see \url{www.astro.umd.edu/$\sim$white/images/eta\_time\_full.html}). }
    \label{fig:ecradio}
  \end{center}
\end{figure}

The radio lightcurve is shown in Figure \ref{fig:ecradio}.  Though the frequency of monitoring of the radio emission is lower than that of the X-ray monitoring, a clear minimum in the radio flux separated by 5.53 years was observed.  Unlike the X-ray emission the radio emission reaches a maximum about halfway through the cycle, then begins a slow decline to minimum.  Also, the radio flux just prior to the radio minimum apparently is declining with each advancing cycle, while the X-ray flux prior to the 2003.5 minimum was substantially brighter than the flux just prior to the 1998.0 minimum.

\subsection{A Colliding Wind Model}

In the colliding wind model  the periodic behavior of the X-ray emission is a consequence of orbital eccentricity and the variation in the amount of absorption in front of the colliding wind shock. 
The X-ray emissivity  and temperature distribution provide constraints on the mass loss rate and terminal velocity of the companion's wind, information which may be impossible to obtain in any other waveband. In turn, this information can be related to the photospheric escape speed using basic radiatively driven wind theory. This helps constrain the mass and radius of the companion.  The photoionizing flux from the companion can be constrained from the free-free emission in the radio. Thus, a nearly complete picture of the identity of the hidden companion may be obtained by detailed modelling of the observed X-ray and radio emission. 

\begin{figure}
  \begin{center}
    \includegraphics[width=\columnwidth]{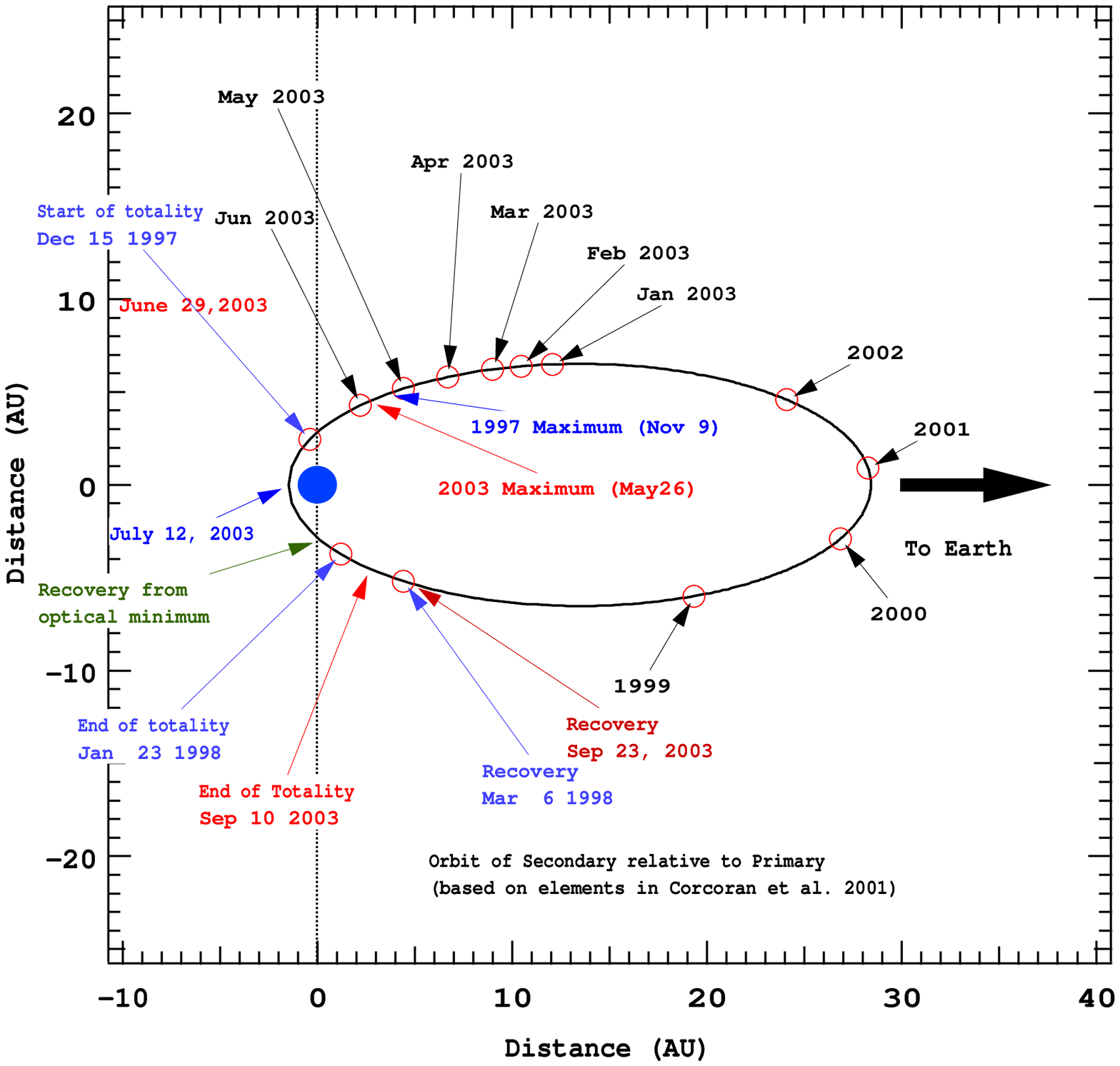}
    \caption{\small Orbit of the companion star around \ec\ based on the orbital elements from \citet{corc01}. }
    \label{fig:orbit}
  \end{center}
\end{figure}

The X-ray spectrum of \ec\ is characterized by a temperature of about 50 million K, which varies little throughout the cycle.  This temperature corresponds to a pre-shock wind velocity of about 3000\ks, which is much higher than the measured wind speed of \ec\ itself, which is only 500\ks \citep{hill01}, so that this emission must arise in the shocked wind of the companion star.  \citet{corc01} compared simple colliding wind models  to the \rxte\ X-ray fluxes.  The orbit implied by this modeling is shown in figure \ref{fig:orbit}.  This orbit is extremely eccentric ($e\approx 0.9$), with the companion behind \ec\ near periastron.   Thus, in this model, the observed X-ray minimum corresponds to a time when the intrinsic X-ray flux is near maximum intensity, but most of this emission is blocked from the observer by the wind from \ec.  The mass loss rates derived from fitting both the emission and the intervening wind absorption is $\dot{M}_{\eta} \approx 10^{-4}$\mdunit\ (which is about a factor of 10 smaller than the mass loss rates attributed to \ec\ based on radio and millimeter observations) while the mass loss rate from the companion is $\dot{M}_{c} \approx 10^{-5}$\mdunit.  The wind characteristics of the companion ($V_{\infty,c}\approx 3000$\ks, $\dot{M}_{c} \approx 10^{-5}$\mdunit) are rather unusual compared to wind parameters derived from analysis of other massive stars; in the study of 24 Galactic O stars by \citet{rep04}, the only star with similar wind parameters is \ec's celestial neighbor HD 93129A (O2If, $V_{\infty}=3200$\ks, $\dot{M}=2.6\times10^{-5}$\mdunit), though HD 93129A itself is a binary which might have biased the derived values.

There are some obvious discrepancies between the observed X-ray emission and the colliding wind models:
\begin{itemize}
\item for constant wind parameters the X-ray flux should be strictly periodic, yet the observed emission shows significant cycle-to-cycle variations, which may indicate evolution of the mass loss from one or both stars;

\item simple 2D models of the wind interaction suggest that the X-ray
emission shoule be symmetric  around periastron  \citep{jmp98}, yet the X-ray brightness prior to eclipse ingress is about a factor of  3 higher than the brightness after recovery from the eclipse. 
This is perhaps indication of enhanced
absorption caused by the distortion of the interaction region by
coriolis forces through periastron passage (which  will require a full 3D
model to determine). Alternatively,  \citet{corc01} suggested that this might be due to the enhancement of the mass loss rate from \ec\ just after periastron passage, perhaps due to some interaction with the companion when the stars are close;

\item unexpected  variations or ``spikes'' occur on a timescale of $\sim 80-100$ days \citep{bish97, corc97, kd98}.   

\end{itemize}

The radio variation is more difficult to model because the radio emission is dependent not only on the photon flux from the companion, but also on the distribution of the circumstellar gas, which is very non-uniform.  The radio flux changes from an unresolved source during the X-ray/radio minimum to a very extended source during the radio maximum, as significant regions of the Homunculus nebula near \ec\ become ionized.  The radio observations help constrain the amount of ionizing radiation from the companion star, which is key in determining the identity  of the companion star.


\section{A (More) Canonical Example: WR140}
\label{wr140}

\begin{figure}
  \begin{center}
    \includegraphics[width=\columnwidth]{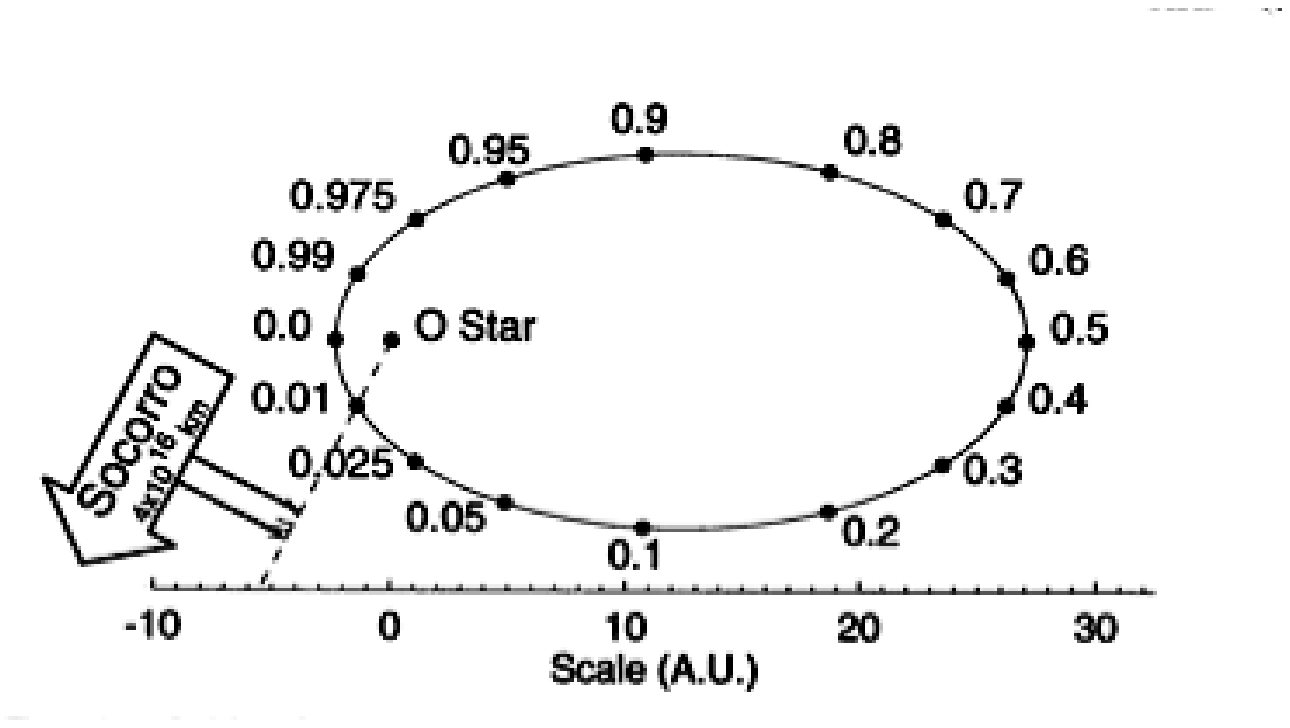}
    \caption{\small Orbit of the WR star around the O4 star adapted from \citet{wb95}.}
    \label{fig:wr140orbit}
  \end{center}
\end{figure}

WR 140 (WC7+O4) is a long-period ($P=2899\pm1.3$ days), eccentric ($e=0.88$) binary \citep{mar03}.  Like \ec\ evidence of this system's  duplicity was first obtained  by the identification of strong, periodic variations in the IR \citep{pw90}.  Unlike \ec, the WR 140 spectrum shows absorptions which arise in the atmosphere of the O4 star so that radial velocity curve solutions exist for both the WR star and the companion.   Thus the orbital elements and mass ratio are well constrained for this binary.  The orbit of the WR star around the O star is shown in Figure \ref{fig:wr140orbit}.

\subsection{X-ray and Radio Emission}

\begin{figure}
  \begin{center}
    \includegraphics[width=\columnwidth]{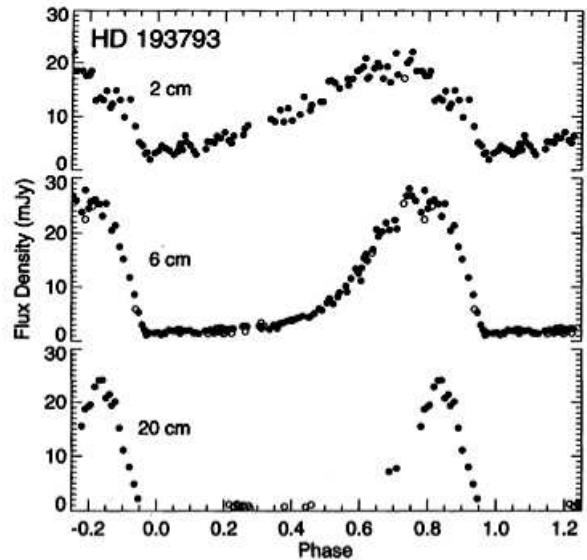}
    \caption{\small Radio observations of the 2, 6, \& 20 cm radio emission from WR 140, adapted from \citet{wb95}, including data from \citet{wb95} and \citet{pw90} , as a function of orbital phase. Phase $=0$ corresponds to periastron passage.}
    \label{fig:wr140radio}
  \end{center}
\end{figure}

WR 140 shows non-thermal radio emission from the motion of  shock-accelerated electrons in a magnetic field near the wind-wind collision shock.  A radio lightcurve for WR 140 through its entire 8-year cycle has been published by \citet{wb95}. They showed that the system exhibits strong variations at 2, 6, and 20 cm (Figure \ref{fig:wr140radio}),  in that the radio emission at all wavelengths is greatest when the O4 companion is in front, and smallest when the O4 star is behind the WR star.  This is presumably due to the the large optical depth at radio wavelengths of the WR star wind, and the smaller radio optical depth of the O star wind.  \citet{wb95} interpreted these wavelength-dependent variations to mean that most of the wind from the WR star is distributed near the star's equatorial plane, noting that it was difficult to explain the observed maximum at 20 cm  when the O star is in front of the WR star, (orbital phase $\phi\approx 0.85$) since the wind from the O star should be opaque at 20 cm.  

\begin{figure}
  \begin{center}
    \includegraphics[width=\columnwidth]{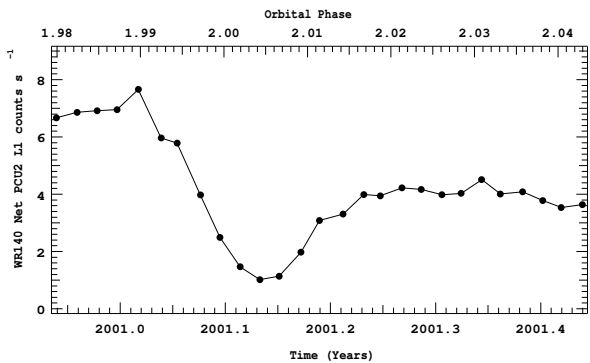}
    \caption{\small \rxte\ observations of the $2-10$ keV X-ray emission of WR 140, adapted from \citet{ap04}. Time in UT years is given along the bottom axis, orbital phase according to the ephemeris of \citet{mar03} along the top. }
    \label{fig:wr140rxte}
  \end{center}
\end{figure}

\begin{figure}
  \begin{center}
    \includegraphics[width=\columnwidth]{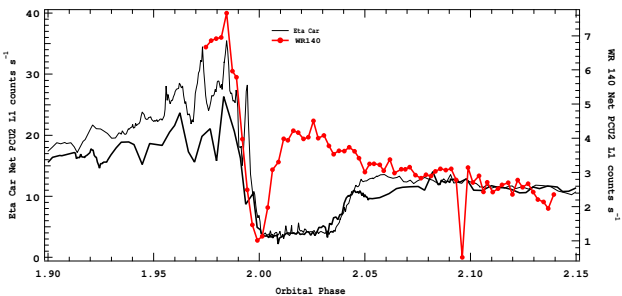}
    \caption{\small Comparison of \rxte\ $2-10$ keV X-ray emission variations of \ec\ (black) and WR 140 (in red). The rate of decline from the X-ray maxima to minima for WR 140 and \ec, and the ratio of the maximum to the recovery level are similar in both stars. Interestingly a ``flare'' seems to occur just before the start of the decline to minimum in both stars. A data dropout is seen  at $\phi\approx 2.09$ in the WR 140 lightcurve.}
    \label{fig:ec_wr140rxte}
  \end{center}
\end{figure}

WR 140 was recognized as an exceptionally bright stellar X-ray source in observations with the \exosat\ satellite \citep{ap87}. Subsequent observations with \rosat\ and \asca\ and later \rxte\ showed that that X-ray emission also varied with orbital phase in a way roughly consistent with simple colliding wind models: the observed emission reached a maximum near periastron passage, while the eclipse by the WR star wind of the X-ray emitting colliding wind shock shortly thereafter produces a brief minimum in the X-ray emission.  The detailed X-ray variation near periastron passage as observed by \rxte\ \citep{ap04} shows a remarkable similarity to the X-ray eclipse of \ec\ (compare similar phases in Figure \ref {fig:lc} and in Figure \ref{fig:wr140rxte}) though in the case of WR 140 the eclipse is quite a bit narrower (see Figure \ref{fig:ec_wr140rxte}.  The X-ray emission is thermal as revealed by \asca\ and \chandra\ spectra though the hot emitting gas is probably out of thermal equilibrium, at least near periastron \citep{ap04}.  

\section{Refining Winds}
\label{winds}

Emission from wind-wind collisions depends upon wind properties so that determination of the level of emission can place severe constraints on mass loss rates and wind speeds.  In addition emission from a wind collision zone can be used as a localized probe of winds and can test how density varies with stellar radius, the smoothness of the stellar wind, and how the wind interacts with the radiation field.

\subsection{Clumping}

As discussed in \S 2, the most direct measure of the stellar mass loss rate is via radio observations of thermal free-free emission.  Mass loss rates derived in this manner may be affected by the presence of non-thermal emission (produced, for example, in the wind-wind collision of unrecognized binaries)  or if the wind is anisotropic or clumped. Clumping means that, rather than being composed of a smooth flow, the wind consists of dense regions immersed within a less dense outflow.  Such clumps may arise due to the inherent doppler instability of radiatively-driven winds as discussed above. Mass loss rates calculated under the assumption that the wind is smooth and spherically symmetric will overestimate the true mass loss rate if the wind is clumped.  Thus understanding how much of the wind is clumped is important to determining the actual amount of mass lost by the star, which plays a key role in determining how the star evolves.

X-ray emission from the wind-wind collision in massive binaries can be used to help determine mass loss rates and the clumping fraction.  Mass loss rates can be derived from the intrinsic luminosity of the X-ray emission which is proportional to \mdot$^{2}$.  The observed emission suffers absorption in the wind of the frontmost star, so that the absorbing column to the X-ray source provides a measure of the density along the line of sight to the X-ray source.  If the orbital geometry is known then the magnitude of the absorbing column to the X-ray source can be used to constrain the mass loss rate of the frontmost star by integrating the density profile through the wind of the frontmost star along the X-ray emitting portion of the wind-wind bow shock.  Mass loss rate estimates from analysis of X-ray emission in colliding wind binaries has been done for $\gamma^{2}$ Velorum \citep{irs96, hs04}, WR 140 \citep{zs00}, and  \ec\ \citep{jmp02b}.  In these cases the mass loss rates derived from the X-ray analyses are lower than those derived from estimates of the
amount of free-free emission by a factor of a few.  For example, \citet{irs96} derived a mass loss rate of $3\times10^{-5}$ \mdunit\ for the WC8 star in $\gamma^{2}$ Velorum based on an observation with the \asca\ X-ray obsevatory when the companion was nearly in front of the WC8 star, a factor of 3 lower than the mass loss rate derived from radio observations.  More recently \citet{hs04}  derived \mdot$\approx2\times10^{-5}$ from analysis of \xmm\ observations when the O star was behind the WC8 primary (assuming the same distance to $\gamma^{2}$ Velorum, $D=450$ pc, used by Stevens et al.). \citet{hs04} showed that the difference derived from the X-ray absorption and the strengths of the X-ray lines imply that about 1/16 of the volume of the WC8 star's wind is clumped.  For \ec, \citet{jmp02b} derived a mass-loss rate from analysis of X-ray spectra which was lower by a factor of a few compared to the rates derived from other methods (like direct measure of free-free emission and spectral synthesis of optical and UV lines), which might be indicative of either clumping, a non-spherical mass outflow, \&/or orbit-related variations in  the mass loss rate.  

\subsection{Radiative Interaction}

In colliding wind binaries the photospheric radiation fields can help moderate the strength of the wind-wind shock.  \citet{sp94} considered the ``radiative inhibition'' of the stellar photons from the companion on the  initial acceleration of the primary's stellar wind, and vice versa.  \citet{og95} and \citet{kg97} considered ``sudden radiative braking'' in which the winds can be almost completely decelerated by the UV flux from the opposing star.  This might help explain why some short-period WR-star binaries like V444 Cygni have unexpectedly low X-ray fluxes.   


\section{Direct Imaging of Colliding Wind Bow Shocks}
\label{imaging}

\begin{figure}
  \begin{center}
    \includegraphics[width=\columnwidth]{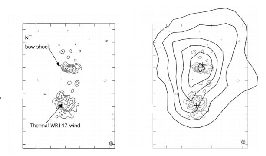}
    \caption{\small \textit{Left}: WR147 imaged by MERLIN \citep{perry97}.  The southern source is resolved thermal emission from the wind of the WR star, while the northern source is non-thermal emission formed in a colliding-wind bow shock.  The crosses mark the positions of the optical stars identified by \citet{virpi98}.  \textit{Right}: \chandra\ HRC X-ray image from \citet{jmp02}.}
    \label{fig:wr147}
  \end{center}
\end{figure}

The ability to resolve sources of emission in colliding wind systems is an important tool  to help understand the physics of the emission mechanisms.  A case in point is the star WR 147, a nearby Wolf-Rayet star which was resolved by VLA \citep{ec92} and MERLIN \citep{perry97} into a southern and northern source separated by 0.6$''$ (Figure \ref{fig:wr147}).  The southern source was identified as thermal free-free emission originating in the wind of the WR star.  Interestingly, at 5 GHz this thermal source is not spherically (or circularly) symmetric.  The northern source shows non-thermal synchrotron emission.  The emission is produced by  electrons which are Fermi accelerated in a colliding wind shock.  \citet{virpi98} identified  an O-type star just behind this shock front, revealing WR 147 as a long-period colliding-wind binary.  

\citet{jmp02} attempted to resolve the colliding wind shock from WR 147 at X-ray energies using the \chandra\ High Resolution Camera, which offers a spatial resolution of about 0.5$''$.  This observation revealed that the X-ray emission is clearly extended but the HRC lacked the spatial and spectral sensitivity needed to clearly distinguish emission from the WR star (or O star) wind from the bow shock.

\section{Finding Binaries}
\label{binaries}

It's important to understand the fraction of massive stars which are binaries in order to understand how massive stars form and evolve.  However conventional means of determining the duplicity of a star can be of limited use in determining whether or not a given massive star has a companion.  Binaries can be identified by periodic radial velocity variations in their photospheric absorption lines, but in massive stars such lines are often rotationally broadened or contaminated by emission or absorption in stellar winds. In systems with large mass ratios the more massive star will show only small-scale radial velocity variations. These problems are especially acute in long-period systems which usually show little radial velocity or photometric variability for most of the orbit.

In principle colliding wind emission can be useful for identifying massive binaries, especially long-period systems where the winds collide at terminal velocities and where the shocks are strong.  \citet{sean00} considered using non-thermal radio emission as a tracer of the massive star binary population.  \citet{sean00} examined 23 WR systems, found 9 non-thermal radio emitters, and showed that 7 of these 9 non-thermal emitters were known to have a spectroscopic or visual binary companion.  They also showed that not all binaries exhibited non-thermal emission, however.  In particular short-period WR binaries tended to be thermal sources, presumably because the non-thermal emitting region is buried within the optically thick WR wind.  

\citet{cher76} suggested using X-ray emission from wind-wind collisions to determine the binary fraction of WR stars.  As is the case with non-thermal radio emission, stellar wind absorption may play a decisive factor in determining how much of the intrinsic emission is observable, especially at energies $E<2$ keV where wind opacity is greatest.  Sources which emit at $E>2$ keV will suffer little absorption for even large column densities ($N_{H}<10^{23}$ cm$^{-2}$), so in this sense X-ray observations have an advantage over radio observations at detecting colliding wind emission.  However the rather coarse spatial resolution of X-ray imaging versus the ability of radio imaging at milli-arcsecond levels is important.  For example  \rosat\  and \asca\ observations \citep{rho00} detected a bright source of hard X-rays associated with HD 164492A, the O7 star which ionizes the Trifid nebula, and suggested that this star might be a colliding wind system.  Subsequent higher spatial resolution observations with \chandra\ \citep{rho03} showed that HD 164492A is actually a soft source and is not overluminous in X-rays compared to single stars, and that the hard emission detected by \rosat\ and \asca\ is from an unrelated nearby object.  Because massive stars usually congregate in associations and clusters, it remains to be seen how many of the catalogued ``X-ray overluminous'' massive stars may be resolved into two or more sources in higher spatial resolution observations.


\section{Conclusions}
\label{conclusions}

Because colliding wind emission can be identified, and the parameters on which it depends (\vterm, \mdot) observationally quantified, observations of colliding wind binary systems at X-ray and radio wavelengths allow for tests of stellar wind structure, and for understanding shock processes and modes of instability.   Colliding winds in massive binaries  provide good laboratories for testing models of shock-generated radio \& X-ray emission.  Comparison of simple analytic models \citep{usov92} and more complex hydrodynamic simulations (Pittard, these proceedings) show that gross characteristics of the emission can be reproduced.  In general, observations of both radio and X-ray observations are necessary to fully test models.  Radio observations provide high spatial resolution imaging to probe the spatial distribution of the shock structure, and can also provide a measure of the magnetic field strength near the shock.  X-ray spectra can provide measures of the spatial distribution of absorbing material in front of the shock, and can be used to measure the flow dynamics by determining the temperature distribution and also via high resolution studies of X-ray emission lines.  

New high resolution X-ray and radio observations however often show complicated behavior which is not yet well matched by the models.  This suggests the importance of realistically modelling the effects of spatial and temporal variations in the stellar winds and Coriolis forces in order to determine the observed emission properties. The complex nature of the wind flows and their apparent inhomogeneities indicates the important need to develop realistic models in 2 \&/or 3-D and the importance of including magnetic fields.


\section*{Acknowledgments}

This work was supported by SAO grant \#GO3-4008A, and this support is most gratefully appreciated.  This research has made use of NASA's Astrophysics Data System. This research has made use of software and data obtained from the High Energy Astrophysics Science Archive Research Center (HEASARC), provided by NASA's Goddard Space Flight Center.  We also thank the organizers for a thought-provoking and interesting meeting.


\begin{thebibliography}{}
\setlength\itemsep{0cm}

\bibitem[Abel, Anninos, Norman, \& Zhang(1998)]{abel98} Abel, 
T., Anninos, P., Norman, M.~L., \& Zhang, Y.\ 1998, \apj, 508, 518 

\bibitem[Abbott, Bieging, Churchwell, \& Cassinelli(1980)]{abcc} Abbott, D.~C., Bieging, J.~H., 
Churchwell, E., \& Cassinelli, J.~P.\ 1980, \apj, 238, 196 

\bibitem[Berghoefer, Schmitt, Danner, \& Cassinelli(1997)]{berg97} Berghoefer, T.~W., Schmitt, 
J.~H.~M.~M., Danner, R., \& Cassinelli, J.~P.\ 1997, \aap, 322, 167 

\bibitem[Bieging, Abbott, \& Churchwell(1989)]{bac89} Bieging, J.~H., Abbott, D.~C., \& Churchwell, E.~B.\ 1989, \apj, 340, 518 

\bibitem[Cassinelli et al.(2001)]{cass01} Cassinelli, J.~P., Miller, N.~A., Waldron, W.~L., MacFarlane, J.~J., \& Cohen, D.~H.\ 2001, 
\apjl, 554, L55 


\bibitem[Castor, Abbott, \& Klein(1975)]{cak} Castor, 
J.~I., Abbott, D.~C., \& Klein, R.~I.\ 1975, \apj, 195, 157 

\bibitem[Castor, McCray, \& Weaver(1975)]{jc75} Castor, J., 
McCray, R., \& Weaver, R.\ 1975, \apjl, 200, L107 

\bibitem[Cherepashchuk(1967)]{cher67} Cherepashchuk, A. M.\ 1967, Peremennye Zvezdy, 16, 226

\bibitem[Cherepashchuk(1976)]{cher76} Cherepashchuk, A.~M.\ 1976, Soviet Astronomy Letters, 2, 138 


\bibitem[Chlebowski, Seward, Swank, \& Szymkowiak(1984)]{chleb84} Chlebowski, T., Seward, F.~D., Swank, J., \& Szymkowiak, A.\ 1984, \apj, 281, 665 

\bibitem[Churchwell et al.(1992)]{ec92} Churchwell, E., 
Bieging, J.~H., van der Hucht, K.~A., Williams, P.~M., Spoelstra, T.~A.~T., 
\& Abbott, D.~C.\ 1992, \apj, 393, 329 

\bibitem[Corcoran, Rawley, Swank, \& Petre(1995)]{corc95} 
Corcoran, M.~F., Rawley, G.~L., Swank, J.~H., \& Petre, R.\ 1995, \apjl, 
445, L121 

\bibitem[Corcoran(1996)]{corc96} Corcoran, M.~F.\ 1996, Revista Mexicana de Astronomia y Astrofisica Conference Series, 5, 54 

\bibitem[Corcoran et al.(1997)]{corc97} Corcoran, M.~F., 
Ishibashi, K., Davidson, K., Swank, J.~H., Petre, R., \& Schmitt, 
J.~H.~M.~M.\ 1997, Nature, 390, 587 


\bibitem[Corcoran, Ishibashi, Swank, \& Petre(2001)]{corc01} Corcoran, M.~F., Ishibashi, K., Swank, J.~H., \& Petre, R.\ 2001, \apj,  547, 1034 

\bibitem[Davidson, Ishibashi, \& Corcoran(1998)]{kd98} 
Davidson, K., Ishibashi, K., \& Corcoran, M.~F.\ 1998, New Astronomy, 3, 
241 

\bibitem[Damineli(1996)]{dam96} Damineli, A.\ 1996, \apjl, 460, L49 

\bibitem[Dorland, Currie, Kaufer, \& Bacciotti(2003)]{dor03} Dorland, B.~N., Currie, D.~G., Kaufer, A., \& Bacciotti, F.\ 2003, Revista Mexicana de Astronomia y Astrofisica Conference Series, 15, 70 

\bibitem[Dougherty \& Williams(2000)]{sean00} Dougherty, S.~M.~\& Williams, P.~M.\ 2000, \mnras, 319, 1005 

\bibitem[Dougherty et al.(2003)]{sean03} Dougherty, S.~M., Pittard, J.~M., Kasian, L., Coker, R.~F., Williams, P.~M., \& Lloyd, H.~M.\ 2003, \aap, 409, 217 

\bibitem[Duncan et al.(1995)]{dunc95} Duncan, R.~A., White, S.~M., Lim, J., Nelson, G.~J., Drake, S.~A., \& Kundu, M.~R.\ 1995, \apjl, 441, L73 

\bibitem[Dyson(1977)]{jd77} Dyson, J.~E.\ 1977, \aap, 59, 161 

\bibitem[Eichler \& Usov(1993)]{eich93} Eichler, D.~\& Usov, 
V.\ 1993, \apj, 402, 271 

\bibitem[Friend \& Abbott(1986)]{fa86} Friend, D.~B.~\& 
Abbott, D.~C.\ 1986, \apj, 311, 701 

\bibitem[Gayley, Owocki, \& Cranmer(1997)]{kg97} Gayley, 
K.~G., Owocki, S.~P., \& Cranmer, S.~R.\ 1997, \apj, 475, 786 

\bibitem[Heger et al.(2003)]{heger03} Heger, A., Fryer, C.~L., 
Woosley, S.~E., Langer, N., \& Hartmann, D.~H.\ 2003, \apj, 591, 288 

\bibitem[Henley, Stevens, \& Pittard(2003)]{dh03} Henley, 
D.~B., Stevens, I.~R., \& Pittard, J.~M.\ 2003, \mnras, 346, 773 

\bibitem[Hillier, Davidson, Ishibashi, \& Gull(2001)]{hill01} 
Hillier, D.~J., Davidson, K., Ishibashi, K., \& Gull, T.\ 2001, \apj, 553, 
837

\bibitem[Howarth \& Prinja(1989)]{hp89} Howarth, I.~D.~\& 
Prinja, R.~K.\ 1989, \apjs, 69, 527 

\bibitem[Ishibashi et al.(1997)]{bish97} Ishibashi, K., 
Davidson, K., Corcoran, M.~F., Swank, J.~H., Petre, R., \& Jahoda, K.\ 
1997, IAU Circ., 6668, 1 

\bibitem[Kahn et al.(2001)]{kahn01} Kahn, S.~M., Leutenegger, M.~A., Cottam, J., Rauw, G., Vreux, J.-M., den Boggende, A.~J.~F., Mewe, R., \& G{\" u}del, M.\ 2001, \aap, 365, L312 


\bibitem[Lucy \& Solomon(1970)]{ls70} Lucy, L.~B.~\& 
Solomon, P.~M.\ 1970, \apj, 159, 879 

\bibitem[Luo, McCray, \& Mac Low(1990)]{luo90} Luo, D., McCray, R., \& Mac Low, M.\ 1990, \apj, 362, 267 

\bibitem[Marchenko et al.(2003)]{mar03} Marchenko, S.~V., et 
al.\ 2003, \apj, 596, 1295 

\bibitem[Morton(1967)]{morton} Morton, D.~C.\ 1967, \apj, 150, 535 

\bibitem[Niemela et al.(1998)]{virpi98} Niemela, V.~S., Shara, 
M.~M., Wallace, D.~J., Zurek, D.~R., \& Moffat, A.~F.~J.\ 1998, \aj, 115, 
2047 

\bibitem[Owocki \& Gayley(1995)]{og95} Owocki, S.~P.~\& 
Gayley, K.~G.\ 1995, \apjl, 454, L145 

\bibitem[Pittard, Stevens, Corcoran, \& 
Ishibashi(1998)]{jmp98} Pittard, J.~M., Stevens, I.~R., 
Corcoran, M.~F., \& Ishibashi, K.\ 1998, \mnras, 299, L5 

\bibitem[Pittard \& Stevens(1997)]{jmp97} Pittard, J.~M.~\& 
Stevens, I.~R.\ 1997, \mnras, 292, 298 

\bibitem[Pittard et al.(2002)]{jmp02} Pittard, J.~M., 
Stevens, I.~R., Williams, P.~M., Pollock, A.~M.~T., Skinner, S.~L., 
Corcoran, M.~F., \& Moffat, A.~F.~J.\ 2002, \aap, 388, 335 

\bibitem[Pittard \& Corcoran(2002)]{jmp02b} Pittard, J.~M.~\& 
Corcoran, M.~F.\ 2002, \aap, 383, 636 

\bibitem[Pollock(1987)]{ap87} Pollock, A.~M.~T.\ 1987, \aap, 
171, 135 


\bibitem[Pollock, et al.(2004)]{ap04} Pollock, A. M. T., et al.\ 2004, ApJ, submitted

\bibitem[Prilutskii \& Usov(1976)]{ps76} Prilutskii, O. F., \& Usov, V. V.\ 1976, Sov. Astron., 20, 2

\bibitem[Repolust, Puls, \& Herrero(2004)]{rep04} Repolust, 
T., Puls, J., \& Herrero, A.\ 2004, \aap, 415, 349 

\bibitem[Rho, Corcoran, Chu, \& Reach(2001)]{rho00} Rho, J., 
Corcoran, M.~F., Chu, Y., \& Reach, W.~T.\ 2001, \apj, 562, 446 

\bibitem[Rho et al.(2003)]{rho03} Rho, J., Ramirez, S., 
Corcoran, M., Hamaguchi, K., \& Lefloch, B.\ 2003, American Astronomical 
Society Meeting, 203,  

\bibitem[Schild et al.(2004)]{hs04}Schild, H., et al.\ 2004, \aap, submitted (astroph/404610)

\bibitem[Stevens, Blondin, \& Pollock(1992)]{sbp} Stevens, 
I.~R., Blondin, J.~M., \& Pollock, A.~M.~T.\ 1992, \apj, 386, 265 

\bibitem[Stevens \& Pollock(1994)]{sp94} Stevens, I.~R.~\& 
Pollock, A.~M.~T.\ 1994, \mnras, 269, 226 

\bibitem[Stevens et al.(1996)]{irs96} Stevens, I.~R., 
Corcoran, M.~F., Willis, A.~J., Skinner, S.~L., Pollock, A.~M.~T., Nagase, 
F., \& Koyama, K.\ 1996, \mnras, 283, 589 

\bibitem[Usov(1992)]{usov92} Usov, V.~V.\ 1992, \apj, 389, 635 

\bibitem[Walborn, Blanco, \& Thackeray(1978)]{nolan78} Walborn, 
N.~R., Blanco, B.~M., \& Thackeray, A.~D.\ 1978, \apj, 219, 498 


\bibitem[Walder(1998)]{rw98} Walder, R.\ 1998, Ap\&SS, 260, 243 

\bibitem[Weaver et al.(1977)]{rw77} Weaver, R., McCray, R., 
Castor, J., Shapiro, P., \& Moore, R.\ 1977, \apj, 218, 377 

\bibitem[Weis, Corcoran, Bomans, \& Davidson(2004)]{kw04} 
Weis, K., Corcoran, M.~F., Bomans, D.~J., \& Davidson, K.\ 2004, \aap, 415, 
595 

\bibitem[White \& Becker(1995)]{wb95} White, R.~L.~\& 
Becker, R.~H.\ 1995, \apj, 451, 352 

\bibitem[Whitelock et al.(1994)]{pw94} Whitelock, P.~A., 
Feast, M.~W., Koen, C., Roberts, G., \& Carter, B.~S.\ 1994, \mnras, 270, 
364 

\bibitem[Williams et al.(1990)]{pw90} Williams, P.~M., van 
der Hucht, K.~A., Pollock, A.~M.~T., Florkowski, D.~R., van der Woerd, H., 
\& Wamsteker, W.~M.\ 1990, \mnras, 243, 662 


\bibitem[Williams et al.(1997)]{perry97} Williams, P.~M., 
Dougherty, S.~M., Davis, R.~J., van der Hucht, K.~A., Bode, M.~F., \& Setia 
Gunawan, D.~Y.~A.\ 1997, \mnras, 289, 10 

\bibitem[Willis, Schild, \& Stevens(1995)]{willis95} Willis,  A.~J., Schild, H., \& Stevens, I.~R.\ 1995, A\&A, 298, 549 

\bibitem[Wright \& Barlow(1975)]{wb} Wright, A.~E.~\& Barlow, M.~J.\ 1975, MNRAS, 170, 41 

\bibitem[Zhekov \& Skinner(2000)]{2000ApJ...538..808Z} Zhekov, S.~A.~\& 
Skinner, S.~L.\ 2000, \apj, 538, 808 


\end{thebibliography}
\end{document}